   \documentclass[12pt,preprint]{aastex}

\newcommand{\snia}{SN~Ia}
\newcommand{\sneia}{SNe~Ia}
\newcommand{\wl}{$\lambda$}
\newcommand{\kms}{\ensuremath{\textrm{km~s}^{-1}}}
\newcommand{\snsd}{SN~2006D}
\newcommand{\co}{C-O} 
\newcommand{\synow}{SYNOW}
\newcommand{\phoenix}{\texttt{PHOENIX}}
\newcommand{\msun}{$M_\odot$}
\newcommand{\bi}{\ensuremath{b_i}}

\newcommand{\discdate}{2006~January~11.2 UTC}
\newcommand{\peakdate}{2006~January~21.8 UTC}

\begin{document}

\title
{
   Nearby Supernova Factory Observations of \snsd:
   On Sporadic Carbon Signatures in Early Type Ia Supernova Spectra
}


\author
{
   The Nearby Supernova Factory \\
   R.~C.~Thomas,\altaffilmark{1}
   G.~Aldering,\altaffilmark{1}
   P.~Antilogus,\altaffilmark{3}
   C.~Aragon,\altaffilmark{1}
   S.~Bailey,\altaffilmark{1}
   C.~Baltay,\altaffilmark{8}
   E.~Baron,\altaffilmark{9}
   A.~Bauer,\altaffilmark{8}
   C.~Buton,\altaffilmark{2}
   S.~Bongard,\altaffilmark{1,5}
   Y.~Copin,\altaffilmark{2}
   E.~Gangler,\altaffilmark{2}
   S.~Gilles,\altaffilmark{3}
   R.~Kessler,\altaffilmark{7}
   S.~Loken,\altaffilmark{1}
   P.~Nugent,\altaffilmark{1}
   R.~Pain,\altaffilmark{3}
   J.~Parrent,\altaffilmark{9}
   E.~P\'econtal,\altaffilmark{4}
   R.~Pereira,\altaffilmark{3} 
   S.~Perlmutter,\altaffilmark{1,6}
   D.~Rabinowitz,\altaffilmark{8}
   G.~Rigaudier,\altaffilmark{4}
   K.~Runge,\altaffilmark{1}
   R.~Scalzo,\altaffilmark{1}
   G.~Smadja,\altaffilmark{2}
   L.~Wang,\altaffilmark{1}
   B.~A.~Weaver\altaffilmark{1,5}
}


\altaffiltext{1}{Physics Division, Lawrence Berkeley National
Laboratory, 1 Cyclotron Road, Berkeley, CA 94720}
\altaffiltext{2}{Institut de Physique Nucl\'eaire de Lyon, UMR5822,
CNRS-IN2P3; Universit\'e Claude Bernard Lyon~1, F-69622 Villeurbanne
France}
\altaffiltext{3}{Laboratoire de Physique Nucl\'eaire et des Hautes
Energies IN2P3 - CNRS - Universit\'es Paris VI et Paris VII, 4 place
Jussieu Tour 33 - Rez de chauss\'ee 75252 Paris Cedex 05}
\altaffiltext{4}{Centre de Recherche Astronomique de Lyon, 9, av.
Charles Andr\'e, 69561 Saint Genis Laval Cedex}
\altaffiltext{5}{University of California, Space Sciences Laboratory,
Berkeley, CA 94720-7450}
\altaffiltext{6}{Department of Physics, University of California,
Berkeley, CA 94720}
\altaffiltext{7}{Kavli Institute for Cosmological Physics, The
University of Chicago, Chicago, IL 60637}
\altaffiltext{8}{Department of Physics, Yale University, New Haven, CT
06250-8121}
\altaffiltext{9}{Homer L. Dodge Department of Physics and Astronomy, 440 W.
Brooks Street, University of Oklahoma, Norman, OK 73019}

\begin{abstract}

We present four spectra of the Type Ia supernova (\snia) 2006D extending
from $-7$ to $+13$ days with respect to $B$-band maximum.  The spectra
include the strongest signature of unburned material at photospheric
velocities observed in a \snia\ to date.  The earliest spectrum exhibits
\ion{C}{2} absorption features below 14,000 \kms, including a
distinctive \ion{C}{2} \wl 6580 absorption feature.  The carbon
signatures dissipate as the SN approaches peak brightness.  In addition
to discussing implications of photospheric-velocity carbon for white
dwarf explosion models, we outline some factors that may influence the
frequency of its detection before and around peak brightness.  Two
effects are explored in this regard, including depopulation of the
\ion{C}{2} optical levels by non-LTE effects, and line-of-sight effects
resulting from a clumpy distribution of unburned material with low
volume-filling factor.

\end{abstract}

\keywords
{
   supernovae: general --- 
   supernovae: individual (\snsd)
}

\section{Introduction}

Type Ia supernovae (\sneia) make valuable standard candles because of
their intrinsic brightness and the homogeneity of their light curves.
Observations of high-redshift \sneia\ are responsible for the recent
revelation that the rate of expansion of the Universe is accelerating
\citep{perl98,garnavich98,riess98,perl99}.  A push to better calibrate
\sneia\ as distance indicators and control systematics in future
high-precision cosmology experiments has helped motivate the search for
deeper physical insight into \snia\ progenitors and their explosion
mechanism.


Certain recent multidimensional hydrodynamical \snia\ explosion models
involve the thermonuclear disruption of a Chandrasekhar-mass \co\ white
dwarf (WD) by \emph{deflagration}, the propagation of a subsonic flame
front through the star \citep[e.g.,][]{gamezo2003,ropke2005}.  A generic
characteristic of these new models is the presence of unprocessed WD
material below the canonical 14,000 \kms\ cutoff seen in W7, a tuned
spherically-symmetric deflagration model \citep{nomoto1984}.  Thus,
carbon detection at such low velocities would generally provide
observational support for the newer multidimensional deflagration
models.  Few \snia\ spectra actually exhibit a robust low-velocity
carbon signature; to date the lowest measured are the \ion{C}{2} \wl
6580 and \wl 7234 features in the spectra of SN~1998aq, which may extend
deeper than 11,000 \kms\ \citep{branch2003}.  The lack of ubiquitous
carbon signatures in the spectra of \sneia, and the observed deficit of
kinetic energy and synthesized nickel indicate that these models are not
viable, though increasingly sophisticated models (e.g., with greater
resolution and more detailed nucleosynthesis) may prove otherwise.  Most
recently, \citet{marion2006} used near-infrared spectroscopy to conclude
that nuclear burning in at least three normal \sneia\ is complete below
18,000 \kms, a conclusion supporting models where an initial
deflagration transitions to a supersonic \emph{detonation}
\citep{khokhlov1991}.

We present Nearby Supernova Factory \citep[SNfactory,][]{aldering2002}
spectroscopy of the \snia\ 2006D, which includes an unambiguous
photospheric-velocity \ion{C}{2} signature.  The data were obtained
using the Supernova Integral Field Spectrograph
\citep[SNIFS,][]{aldering2002,lantz2004} on the University of Hawaii
2.2-meter telescope on Mauna Kea.  Our focus here is on the early
photospheric-phase spectra of \snsd; an analysis involving all of our
data will appear elsewhere.

\section{Data \& Analysis}

\begin{figure}
   \centering
   \includegraphics[scale=1.00,clip=true]{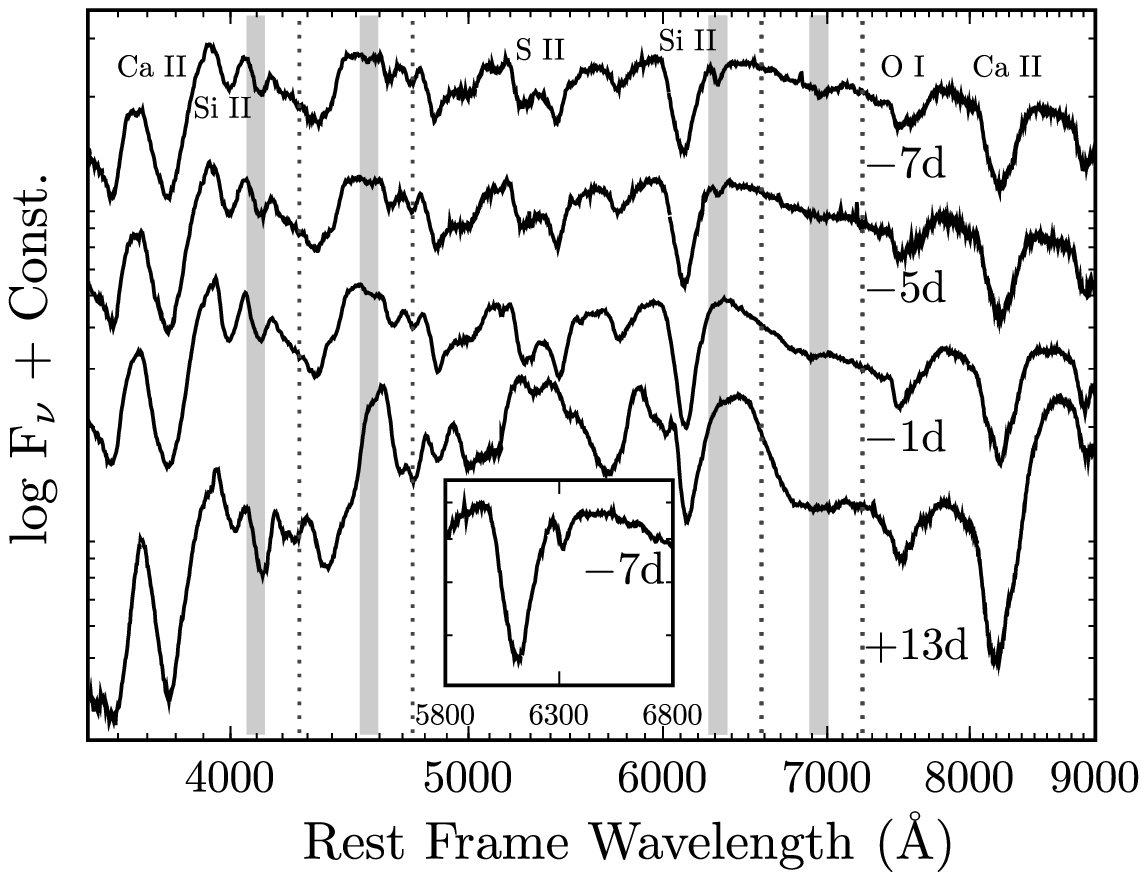} 
   \caption{SNIFS spectroscopy of \snsd.  Phases are expressed relative
   to a $B$-band peak brightness date of \peakdate.  Dotted vertical
   lines mark rest wavelengths of \ion{C}{2} lines \wl\wl 4267, 4745,
   6580, and 7234.  Dark bands indicate blueshifts between 10,000 and
   14,000 \kms\ with respect to these lines, typical of the velocity at
   the photosphere at these phases.  The inset is a zoom of the region
   around the 6320~\AA\ notch in the $-7$ day spectrum.}
   \label{fig:time_series}
\end{figure}

The Brazilian Supernova Search \citep{colesanti2006} discovered \snsd\
in MCG -01-33-34 \citep[$z = 0.00853$,][]{davoust2004} on \discdate.
Using a SNIFS spectrum obtained January~14.6~UTC, we classified \snsd\
as a \snia\ one week prior to maximum brightness \citep{atel}.  This and
three subsequent spectra appear in Figure~\ref{fig:time_series} and are
summarized in Table~\ref{tab:spectra}.  The spectra were reduced using
our dedicated data-reduction procedure, similar to that presented in
\S~4 of \citet{bacon2001} and \S~2 of \citet{aldering2006}.  No
correction for interstellar reddening has been applied.  The
intermediate-mass element (IME) spectral features typical of premaximum
\sneia\ are labelled on the first spectrum for reference.
A preliminary light curve fit derived from SNIFS acquisition
images suggests a $B$-band peak date of \peakdate.  The light curve is
more narrow than is typical for a spectroscopically normal \snia\ ---
the derived time-scale ``stretch'' parameter $s \sim 0.75$ is
significantly smaller than in the normal case ($s \equiv 1$).  This
value is similar to that derived from the light curves of SNe~1986G and
1992bo \citep[0.74 and 0.73 respectively,][]{guy2005}, but not as
extreme as in the sub-luminous SNe~1991bg or 1999by \citep[$s \sim
0.6$;][]{ruiz2004, garnavich2004,jha2006}.

The dotted vertical lines in Figure~\ref{fig:time_series} indicate the
rest-frame wavelengths of the four \ion{C}{2} lines that would be the
strongest under the assumption of local thermodynamic equilibrium (LTE)
at 10,000~K.  The shaded regions to the blue of each line cover Doppler
shifts between 14,000 and 10,000 \kms, representing velocities typical
of the photosphere before and around peak brightness.  A striking
feature of the $-7$ day spectrum is the $\vee$-shaped absorption
centered at 6320~\AA\ (rest frame), which we attribute to \ion{C}{2} \wl
6580 at a Doppler shift of 12,000 \kms.  In the first spectrum, two much
weaker notches fall within the bands of the shaded regions corresponding
to \ion{C}{2} \wl\wl 4745 and 7234.  These two features and the
6580~\AA\ notch weaken by day $+13$.

Another interesting feature in our spectra is the notch centered at
4100~\AA, just to the red of the small absorption typically attributed
to \ion{Si}{2} \wl\wl 4128,4131.  If the 4100~\AA\ absorption is due to
\ion{C}{2} \wl 4267, then it extends from 14,000 \kms\ down to 8,000
\kms, slower than the red edge of the \wl 6580 absorption by 2,000 \kms\
(of course, such a low velocity edge could be the result of
line-blending).  While the other \ion{C}{2} features dissipate by day
$+13$, the 4100~\AA\ notch persists and the absorption along its red
edge strengthens somewhat.  In later spectra not shown here, this
feature steadily becomes washed out by iron peak lines in this part of
the spectrum.  If this feature does arise from \ion{C}{2} \wl 4267, then
it does so nonthermally --- in LTE at 10,000 K it would be approximately
30 times weaker than \ion{C}{2} \wl 6580; we return to this point
shortly.

\begin{figure}
   \centering
   \includegraphics[width=1.00\textwidth,clip=true]{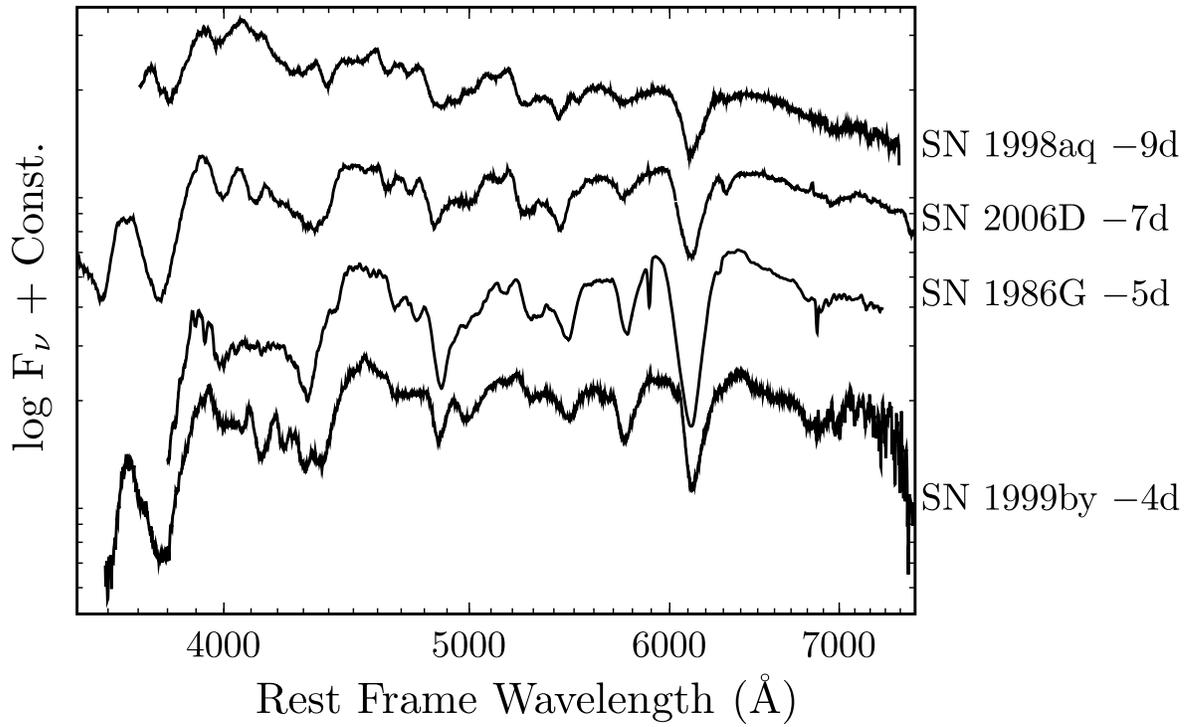} 
   \caption{Comparison of the $-7$ day spectrum of \snsd\ to premaximum
   spectra of SNe~1998aq, 1986G, and 1999by.}
   \label{fig:spec_comp_early}
\end{figure}

In Figure~\ref{fig:spec_comp_early}, we plot our earliest spectrum
together with premaximum spectra of SNe~1998aq, 1986G, and 1999by.  The
spectrum of SN~1998aq is an excellent match to that of \snsd,
particularly redward of 4400~\AA, aside from the stronger \ion{C}{2} \wl
6580 absorption in the latter.  Blueward of 4400~\AA, the agreement is
not as good; the 4100~\AA\ notch in the spectrum of \snsd\ has no
counterpart in that of SN~1998aq save for a minor wiggle \citep[in fact,
this wiggle has been previously observed in the spectra of \sneia, see
Figure 2 of][]{branch2006}.  The spectra of SNe~1986G and 1999by exhibit
larger \ion{Si}{2} 5800~\AA\ to \ion{Si}{2} 6150~\AA\ absorption ratios
than does \snsd, indicative of lower temperature than in more
spectroscopically normal \sneia\ \citep{nugent1995}.  The region from
4000 to 4400~\AA\ in these two spectra is dominated by \ion{Ti}{2}, also
characteristic of lower temperature.  The differences between the
spectra of \snsd\ and SN~1998aq in this region, and the possibility
based on its light curve that \snsd\ is slightly sub-luminous, suggest
that \ion{Ti}{2} may play a role in the formation of its spectrum.

\begin{figure}
   \centering
   \includegraphics[width=1.00\textwidth,clip=true]{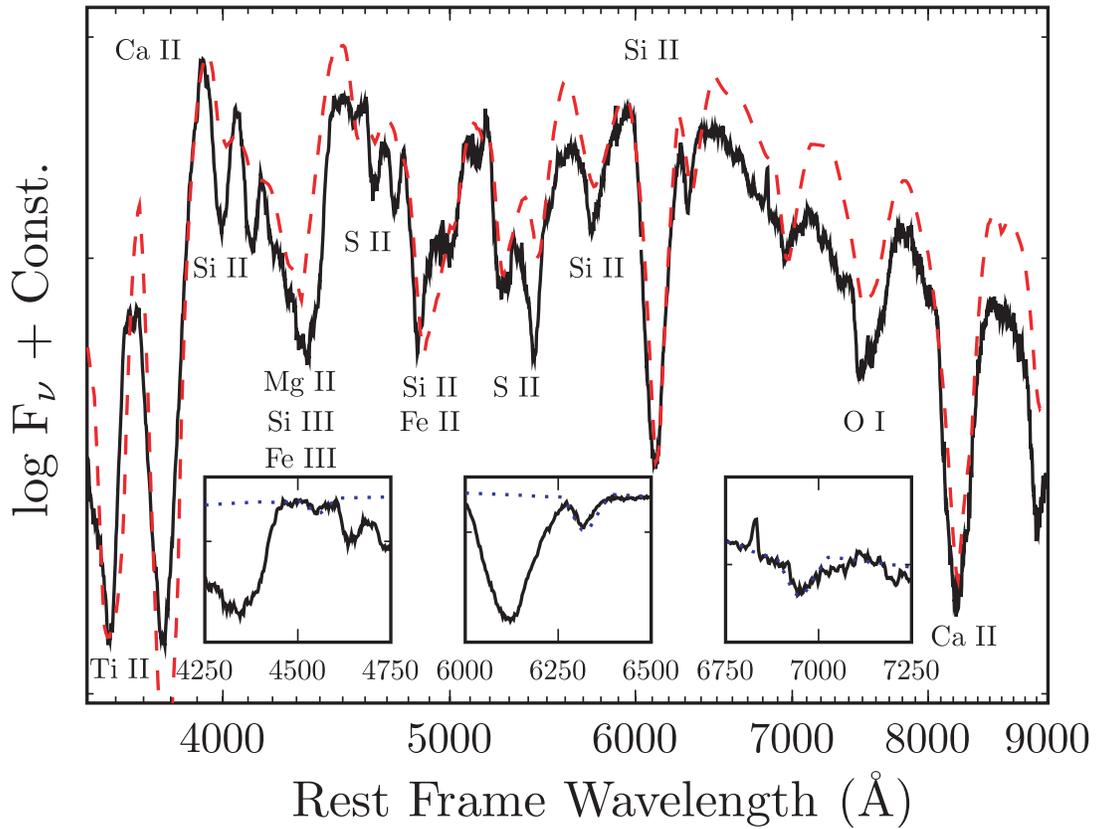} 
   \caption{Synthetic spectra overlaid on the $-7$ day spectrum of
   \snsd.  Inset plots show the coincidence between the synthetic
   \ion{C}{2} spectrum and the data.}
   \label{fig:synow_m10}
\end{figure}

More detailed SN spectral line identification requires an accounting for
line blending, so we compare the $-7$ day spectrum of \snsd\ with spectra
generated by the highly-parameterized SN spectrum synthesis code \synow\
\citep[e.g.,][]{branch2003}.  \synow\ uses a simplified model of a SN
atmosphere consisting of a sharply-defined continuum-emitting core
surrounded by a line-forming region.  Line formation is treated under
the Sobolev approximation, with line opacity parameterized as a function
of ejection velocity and with relative strengths for a given ion set by
assuming Boltzmann-factor population of the levels.  A synthetic
spectrum is overlaid on the observed $-7$ day spectrum in
Figure~\ref{fig:synow_m10}.

The dashed fit reproduces most of the main absorption features of the
spectrum, including those of IME ions \ion{Ca}{2}, \ion{Si}{2}, and
\ion{S}{2}.  A velocity at the photosphere of 11,000~\kms\ produces
satisfactory results.  The blend at 4700~\AA\ consists of lines of
\ion{Fe}{2} and \ion{Si}{2}, while the blend from 4100~\AA\ to 4400~\AA\
consists of lines from \ion{Fe}{3}, \ion{Mg}{2}, and \ion{Si}{3}, and
indeed some \ion{Ti}{2} (which also fits the absorption at 3600~\AA\ as
well).  This \ion{Ti}{2} detection is consistent with cooler,
fast-declining
\sneia.

The sharpness of the \wl 6320 feature, and the lack of a
discernable accompanying emission feature to its red motivates us to
distribute \ion{C}{2} Sobolev line opacity in a layer ``detached'' from
the SYNOW photosphere \citep{jeffery1990}.  An absorption feature
without a corresponding P~Cygni emission may also be indicative of a
clump along the line of sight partially obscuring the photosphere
\citep{thomas2002}.  A thin shell of Sobolev line opacity immediately
above the photosphere would result in a flat absorption trough instead
of a sharper downward spike.  In the synthetic spectrum, \ion{C}{2}
opacity is detached from the photosphere to a velocity of 12,000 \kms,
producing a good match to the \ion{C}{2} \wl\wl 6350 and 7234 features.
The dotted synthetic spectra in the insets of Figure~\ref{fig:synow_m10}
are from a pure \ion{C}{2} spectrum, overlaid on the neighborhoods of
the observed \ion{C}{2} \wl\wl 4745, 6580, and 7234 features.

We can account for the 4100~\AA\ feature with a separate higher velocity
\ion{Ti}{2} component (included in the fit with $v > 17,000$ \kms) but
ultimately reject this hypothesis due to the lack of any sign of
high-velocity components to explain other spectral features.  In
particular, \ion{Ca}{2} opacity should trace the same high-velocity
processed material, and we detect no such signature.

In the absence of any obvious alternative to \ion{Ti}{2} for the
4100~\AA\ notch, we suggest that the feature could come from \ion{C}{2}
\wl 4267, stimulated by some non-LTE process.  To investigate this
hypothesis, we examined the non-LTE departure coefficients ($b_i \equiv
n_i / n_i^{LTE}$, where $n_i$ is the population number of the atomic
level, $i$) of the lower levels of the optical \ion{C}{2} lines in a
\phoenix\ calculation of the model W7 \citep{baron2006}.  In the
\co-rich zone, departure coefficients in optical lines are found to be
significantly smaller than unity ($10^{-7}$ to $10^{-3}$).  Furthermore,
the ratio of the \wl 4267 lower-level \bi\ to that of \wl 6580 is
observed to range between 1 and 60.  This situation arises due
to more efficient recombination through the ultraviolet resonance line
\ion{C}{2} \wl 687, so that recombination into the optical levels is
suppressed and they take on level populations significantly smaller than
LTE.  The size of this ratio could account for the persistence of
\ion{C}{2} \wl 4267 while the other \ion{C}{2} lines dissipate.  Further
detailed models are required to discover whether this effect is indeed
generic and holds in other models besides W7.  This effect may also
resolve the long-standing difficulty of making a reliable identification
of \ion{C}{2} in the spectra of \sneia.

\section{Discussion}

That some \snia\ spectra possess \ion{C}{2} signatures, particularly
before maximum light, is now quite clear.  Past tentative
identifications of \ion{C}{2} \wl 6580 in such cases as SN~1990N
\citep{mazzali2001} and SN~1999ac \citep{garavini2005}, and previously
unidentified notches seen in SN~1994D and SN~1996X pointed out by
\citet{branch2003} now seem more meaningful.  These detections and their
circumstances prompt consideration of several issues relevant to the
study of \snia\ explosion models.


The primary issue is the amount of carbon detected.  To place a lower
limit on the carbon mass (particularly in light of the small \bi's
observed in the \phoenix\ model), we compute LTE electron level
populations and corresponding \ion{C}{2} \wl 6580 Sobolev optical depth
on a temperature-density grid, assuming a \co-rich composition.  The
absolute minimum density giving rise to optical depth $\tau \sim 1$ at
10 days after outburst is $5.9 \times 10^{-15}$~g~cm$^{-3}$.  Treating
this as an average density results in a lower limit of 0.007 \msun\ of
carbon (0.014 \msun\ of unburned material) over the velocity range
10,000 to 14,000 \kms\ where the absorption feature is detected.  This
value is smaller than the 0.085 \msun\ of unburned material present at
the same velocity interval in the multidimensional model \emph{b150\_3d}
of \citet{ropke2006}.  

If the behavior of the \ion{C}{2} \bi's seen in the \co-rich zone of the
W7 \phoenix\ model is generic, then the amount of unburned material
detected could be \emph{significantly higher} than this estimate.  The
\bi's need not be so miniscule as observed in the \phoenix\ model ---
here only a factor of 6 is needed to produce general agreement with
model \emph{b150\_3d}.  Extremely small departure coefficients drive the
mass up but the mass is obviously limited by the mass-energy budget of
the progenitor, fixed by the Chandrasekhar mass.  One clear way to make
the detection consistent with this constraint is for the unburned
material to be distributed in clumps if the \bi's are indeed very small.

The presence of unburned material at the observed velocities may appear
more consistent with published multidimensional deflagration models than
with delayed-detonation models, but the situation is not quite so
simple.  For example, \citet{hoflich2002} demonstrate that sub-luminous
delayed-detonation models can fit the light curve and infrared spectra
of the sub-luminous SN~1999by, and account for observed \ion{C}{1}
features observed with a small amount of carbon extending below 14,000
\kms.  Interestingly, the best-fit model for SN~1999by is from a grid of
delayed-detonation models with variable transition density, and
corresponds to the lowest luminosity/lowest transition density in the
grid.  The trend in those models seems to indicate that less luminous
models, with lower transition densities, have a correspondingly
decreased carbon cutoff velocity.  Thus, if the more luminous \snsd\
were to fall on that particular grid of models, it would need to
correspond to a higher transition density and have a higher carbon
cutoff velocity than SN~1999by, at odds with our observations.  More
speculatively, \citet{gamezo2004} outline a scenario in which a small
amount of unburned material in pockets could be left by a
multidimensional delayed detonation --- as the detonation wave processes
material left between plumes generated during the deflagration phase,
abrupt twists and turns could cut it off from fresh WD fuel.
Elucidating the details (the mass and spatial distribution of such
pockets) will rely on progress in future multidimensional
delayed-detonation models.

\begin{figure}
   \centering
   \includegraphics[width=0.4\textwidth,clip=true]{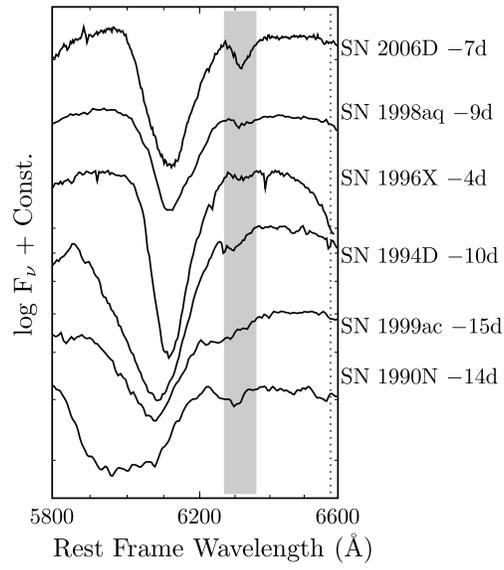} 
   \caption{Comparison of \ion{C}{2} \wl 6580 feature across several
   premaximum \snia\ spectra.  The spectra are resampled into 10~\AA\
   bins.}
   \label{fig:c_comp}
\end{figure}

It is interesting to consider the significance of the sporadic and
diverse nature of the detections of the \ion{C}{2} \wl 6580 feature
itself.  The premaximum rest-frame 5800-6600~\AA\ spectra of \snsd\ and
five other \sneia\ plotted in Figure~\ref{fig:c_comp} exhibit a variety
of strengths and velocity ranges in \ion{C}{2} \wl 6580.  The
theoretical multidimensional models themselves may suggest an
explanation for this phenomenon beyond any non-LTE effect --- the
distribution of unburned material at these velocities could be quite
asymmetric, confined to clumps of small filling factor compared to the
photosphere.  Thus, differences from SN to SN would naturally be
explained as a function of the distribution of the clumps with respect
to the observer's line of sight.  This argument is consistent with the
complementary finding by \citet{thomas2002}, where the relatively
consistent shape and depth of \ion{Si}{2} \wl 6355 absorption at maximum
light indicates that the distribution of IME's in \sneia\ cannot be
highly asymmetric.  Of course, differences in the details of temperature
structure and ionization from one SN to another may be in play, though
we would expect temperature effects to run along a continuum correlated
to, say, radioactive nickel mass.  If low-velocity carbon detections
are sporadic and less correlated with luminosity \citep[note that the
light curve of SN~1998aq is characterized as ``typical,''][]{branch2003},
one would expect the geometrical argument to be at least as plausible.
An underlying mechanism for either explanation (or both) will depend on
further observations of the earliest \snia\ spectra that probe the outer
layers of these events, and on improved detailed modeling of these
spectra.

\section{Conclusions}

We have presented the strongest evidence to date of unburned ejecta at
low velocity in the early optical spectra of a \snia.  The case of \snsd\
reiterates the importance of early-time spectroscopy (and
spectropolarimetry) for \snia\ research.  Complimentary observations in
the nebular phase (one year after explosion) may present further
opportunities to probe the distribution and amount of unburned material
at low velocity in \snsd;  while their predictions are not strictly
applicable at the velocities considered in this work, the
multidimensional spectrum synthesis models of \citet{kozma2005} suggest
that unburned material at velocities below 10,000 \kms\ roughly
consistent with our mass estimates may be detectable at late times as
forbidden [\ion{O}{1}] and [\ion{C}{1}] emission.  

\acknowledgments

The authors thank the anonymous referee for many helpful comments.
We are grateful to the technical and scientific staff of the University
of Hawaii 2.2-meter telescope for their assistance in obtaining these
data.  The authors wish to recognize and acknowledge the very
significant cultural role and reverence that the summit of Mauna Kea has
always had within the indigenous Hawaiian community.  We are most
fortunate to have the opportunity to conduct observations from this
mountain.  We also thank F.~R\"opke for kindly providing the
angle-averaged mass distribution of model \emph{b150\_3d}.  Our analysis
included data from the University of Oklahoma Online SUpernova SPECTrum
(SUSPECT) Archive.  This work was supported in part by the Director,
Office of Science, Office of High Energy and Nuclear Physics, of the
U.S. Department of Energy under Contract No. DE-FG02-92ER40704, by a
grant from the Gordon \& Betty Moore Foundation, by  National Science
Foundation Grant Number AST-0407297, and in France by support from
CNRS/IN2P3, CNRS/INSU and PNC.  This research used resources of the
National Energy Research Scientific Computing Center, which is supported
by the Office of Science of the U.S.  Department of Energy under
Contract No. DE-AC02-05CH11231.  We also acknowledge support from the
U.S. Department of Energy Scientific Discovery through Advanced
Computing program under Contract No.  DE-FG02-06ER06-04.

{\it Facilities:} 
\facility{UH:2.2m (SNIFS)}

\clearpage

\begin{deluxetable}{llccccl}
   \tabletypesize{\scriptsize}
   \tablecaption{Journal of Spectroscopy of \snsd\label{tab:spectra}}
   \tablewidth{0pt}
   \tablehead
   {
      \colhead{Julian Day}               &
      \colhead{UTC Date}                 &
      \colhead{Exposure}                 &
      \colhead{Airmass}                  &
      \multicolumn{2}{c}{Median S/N\tablenotemark{a}} &
      \colhead{Conditions}               \\
      \colhead{}                         &
      \colhead{(2006)}                   &
      \colhead{(s)}                      &
      \colhead{}                         &
      \colhead{blue}                     &
      \colhead{red}                      &
      \colhead{}      
   }
   \startdata
   2453750.10 & Jan.\ 14.60 &  900 & 1.20 & 76  & 85  & non-photometric \\
   2453752.12 & Jan.\ 16.62 & 1000 & 1.16 & 72  & 69  & cloudy          \\
   2453756.07 & Jan.\ 20.57 & 1800 & 1.22 & 205 & 136 & near Moon       \\
   2453770.11 & Feb.\ 03.61 & 1000 & 1.16 & 72  & 93  & photometric     \\
   \enddata
   \tablenotetext{a}{
   Blue channel: 3500-5050 \AA\ with 2.4 \AA\ bins. Red channel:
   5000-9000 \AA\ with 3.0 \AA\ bins.}
\end{deluxetable}

\clearpage

\clearpage

\clearpage

\clearpage

\end{document}